# An Examination of Rotational Effects on Large Wind Turbine Blades

Galih S.T.A. Bangga, Thorsten Lutz, Ewald Krämer
*Institute of Aerodynamics and Gas Dynamics, University of Stuttgart*
*Pfaffenwaldring 21, 70569, Stuttgart, Germany*
bangga@iag.uni-stuttgart.de

CFD studies have been performed to demonstrate the effects of rotation on large wind turbine blades. The 10 MW blade of the Advanced Aerodynamic Tools for Large Rotors (AVATAR) project was selected in the present investigation. In order to demonstrate the rotational augmentation, lift coefficients of the blade sections were compared with the two-dimensional simulations at consistent inflow conditions. The increase of the lift coefficient was observed at the inboard section of the blade, showing the effect of rotation. A significant delay of separation point as well as a massive reduction of the wake size occur at the inboard blade section. The results of the present manuscript provide more information of the rotational augmentation occurring in the near hub region of wind turbine blades.

## I. INTRODUCTION

The accurate prediction of wind turbine performance remains challenging due to the flow complexity at the inner part of the blades. The difficulties come from the fact that rotation plays an important role on the boundary layer development under highly separated flow at the post-stall regime. The detailed mechanism of the flow physics is however far from being well understood [1-3], hindering most of blade element momentum (BEM) models. For instance, the BEM calculations underpredicted the power of the NREL Combined Experiment (Phase II) turbine by approximately 15-20% [4].

The effects of rotation on the rotating blade was introduced by Himmelskamp [5], showing that the lift enhances and the stall is delayed compared with the two-dimensional (2D) airfoil data. It is believed that the centrifugal and the Coriolis forces have a huge contribution regarding this matter. The centrifugal force, which has a strong influence on the separated area [6], transports the flow in the radial direction from the root towards the middle region of the blade, reducing the pressure level on the suction side of the blade. Afterwards, the Coriolis force produces favourable conditions in the boundary layer development, mitigating the separation line along the span towards the trailing edge [7].

Most investigations regarding the underlying phenomena were focused on small stall-controlled wind turbines. The wind turbine size increases significantly nowadays, and rotational effects are expected to be less important for these large wind turbine blades [1]. However, it should be kept in mind that the tip speed ratio of larger turbines is comparable with the smaller one [2], resulting in the congruous value of Rossby number, and this leads to the similar effects of rotation as consequence. Currently, information available on this matter is inadequate and deeper investigations are necessary. Therefore, the present works are addressed specifically to investigate the rotational effects on the large wind turbine blades. The sectional lift coefficients along the blade radius are compared with the two-dimensional (2D) simulations to demonstrate the rotational effects. The analysis of the flow field at the inboard blade section is performed to show the influence of the rotation on separation point as well as the wake development. Comments are made in the flow physics of the blade under rotational motion.

## II. METHODS

### A. The Wind Turbine

The calculations were performed on the 10 MW blade of the Advanced Aerodynamic Tools for Large Rotors (AVATAR) project [8,9]. The calculations were conducted at the wind and rotational speeds respectively $U = 10.5$ m/s and $n = 9.0218$ rpm, and the pitch angle was 0.0°. A uniform inflow condition has been selected to isolate the rotational effects from unsteady phenomena, e.g. dynamic stall.

As initialization, an unsteady calculation was performed until it reached periodic convergence of power and thrust (the condition where the mean value of power and thrust does not change over the azimuth angles). It was achieved after 15 revolutions. Then, additional two revolutions were simulated to extract the data. The standard averaging procedure was applied for the last two revolution. All presented data in this manuscript are averaged values, unless otherwise stated.

### B. The CFD Setup

The unsteady numerical investigations of the flow over rotating blade have been conducted by utilizing the computational fluid dynamics (CFD) code *FLOWer*. The (URANS) *SST-k$\omega$* turbulence model was employed as it provides reasonable accuracy at high adverse pressure gradient flows [6,10-15]. Fully turbulent scheme was chosen in the present study since the objective of the investigation is to demonstrate the rotational augmentation – not to predict the accurate performance of the turbine. Dual time-stepping was utilized to obtain second order accuracy in time. The timestep value is 3°. The calculations have been performed at smaller timesteps, and there were no significant change in the resulting averaged power, thrust, and sectional loads.

The blade mesh was C-type and constructed using an automatic grid generator script developed at the institute [16] by employing the grid generator Gridgen. Fig. 1 shows the mesh of the blade. The meshes of the other structures were built by hand using Pointwise. The Chimera interpolation method was employed in the overlapping grid areas. The meshes were build 120-degrees symmetry of the rotor, assuming periodicity of the flow by using periodic boundary condition; therefore, only one blade is simulated, reducing the computational cost. The size of the background domain is at least 6 and 14 blade radius in length and background radius, respectively. The farfield boundary condition was applied in the background domain.

To show that the solutions are independent of the spatial resolution, grid convergence index (*GCI*) studies according to Celik *et al.* [17] have been performed. The calculations were performed using three different blade meshes. All meshes have 280 and 128 cells respectively along the sectional airfoil and the normal direction. The coarse mesh has 136 cells in the spanwise direction, and this corresponds to the total number of blade mesh of 8.1 x $10^6$ cells. The spanwise cells number was refined with the refinement factor around 1.4, resulting in 280 cells (10.9 x $10^6$ cells) for the medium mesh and 272 cells (15.9 x 106 cells) for the fine mesh. The background, hub and nacelle, and shaft grids have 4.38 x $10^6$, 0.69 x $10^6$ and 1.08 x $10^6$ cells, respectively. They were maintained constant without refinement.

To demonstrate the rotational augmentations, the three-dimensional (3D) results need to be compared with the 2D calculations of the blade sections. The calculations were performed at the same inflow conditions, for instance, in Reynolds number (*Re*), Mach number (*Ma*), and also the angle of attack (*AoA*). The extraction of the angles of attack is based on the reduced axial velocity method described in Ref. [18]. The CFD setup and the mesh quality for the 2D case was identical with the 3D simulations. For comparison, the 2D calculations were also performed using the boundary layer code *XFOIL*.

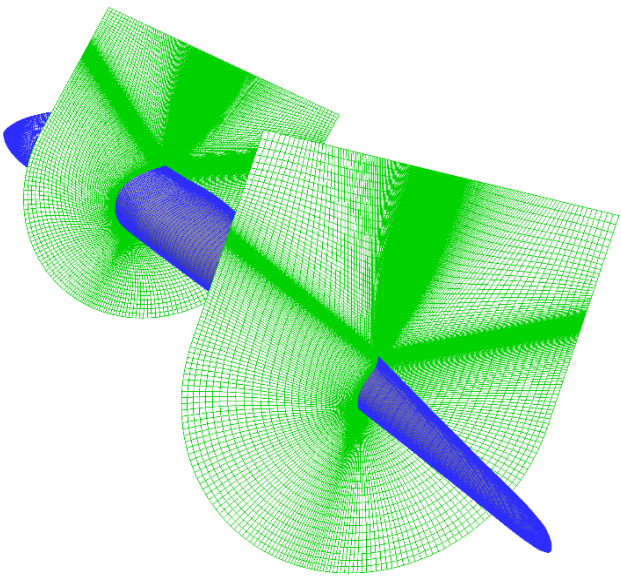

Fig. 1 Mesh of the blade used in the calculations.

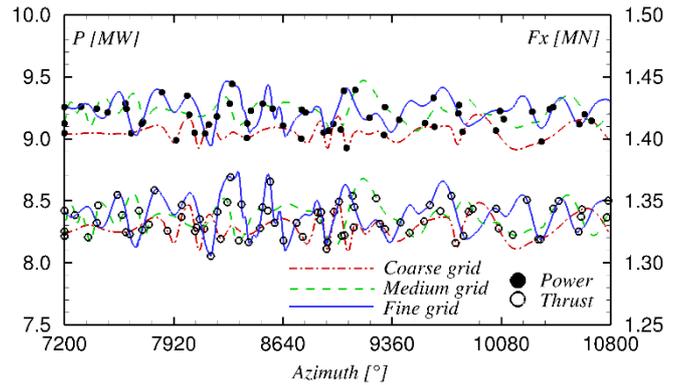

Fig. 2 Power and thrust histories computed using three different meshes.

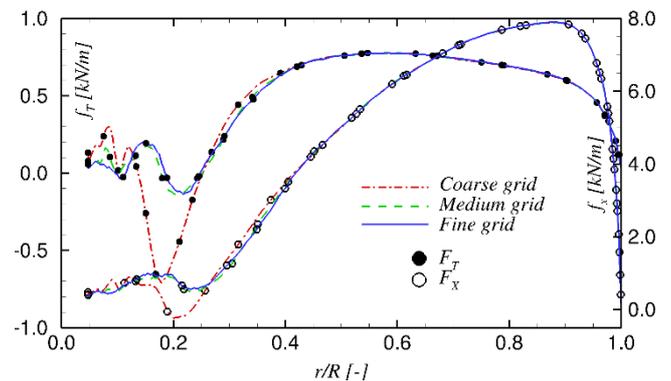

Fig. 3 Sectional forces computed using three different meshes.

### III. RESULTS AND DISCUSSION

#### A. Spatial Discretization Study

Fig. 2 shows the plot of total power (*P*) and thrust ($F_X$) histories over the azimuth. The calculations were undertaken using three different grids as mentioned in the previous section. It can be seen that the averaged values of those parameters are very close each other. However, it seems that the coarse grid produces lower value in the both parameters compared with the other grids. This is the results of sectional forces underprediction at the inboard section of the blade, see Fig. 3. The massive underpredictions of the sectional tangential ($f_T$) and thrust forces ($f_X$) occur at about 20% radius using the coarse mesh, while the values of predicted $f_T$ and $f_X$ using the medium and fine meshes are almost undistinguishable. Moreover, the *GCI* values of the present study are 0.003% and 0.0497% in terms of power and thrust, respectively. This shows that the grid is spatially converged. Therefore, considering the computational cost and accuracy, the medium grid was chosen in all simulations presented in the manuscript.

It was observed that the number of cells of the coarse mesh does not fine enough to resolve the flow field at the inboard section. This comes from the fact that the inboard section produces severely unsteady flow due to separation, and this cannot be accurately predicted by the coarse mesh. It can be seen in Fig. 4 that the surface pressure coefficient ($C_p$) distribution predicted by the coarse mesh behaves in different manner compared with the other meshes. At the inboard blade sections, for *r/R* < 0.1, it seems that the coarse mesh calculated smaller separation, which results in the higher tangential force in Fig. 3. However, an interesting phenomenon can be seen at $0.1 < r/R < 0.3$ where earlier separation line is shown. This results in the massive underpredictive values of $f_T$ and $f_X$. It is also worthwhile to note that the prediction of separation position is a quite problematic issue – even for nowadays computational methods. Moreover, the turbulence is not resolved, but is modelled. Further examinations are necessary to study this matter as a strong radial flow occurs.

*B. Validation*

The CFD setup of the present studies are validated by comparing the resulting total power and thrust with the numerical results from the other project partners [9]. Description of the other codes and numerical approaches were given in Ref. [9]. Fig. 5 shows the validation of the *FLOWer* results against 7 different codes. *FLOWer* predicted very similar power value with *DTU EllipSys3D*, *DTU Miras*, and *NTUA MaPFlow*, while the other codes predicted higher values. Furthermore, predicted thrust of *DTU EllipSys3D* is almost exactly the same value with the present works; therefore, this validates the present CFD setup.

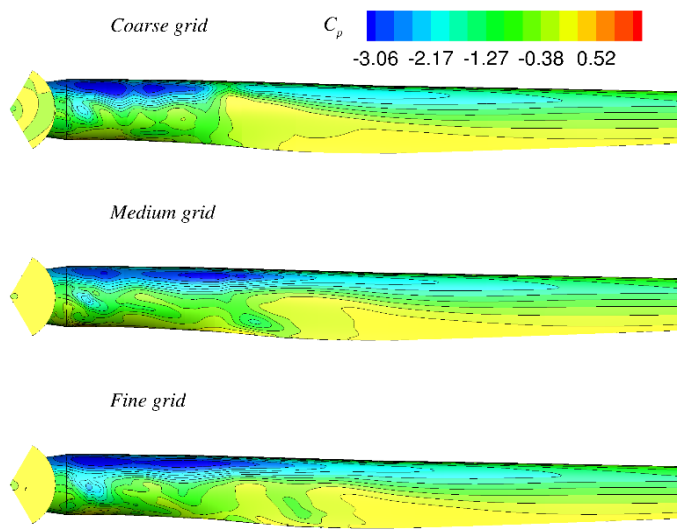

Fig. 4 Instantaneous $C_p$ distribution on the suction side at azimuth 10800°.

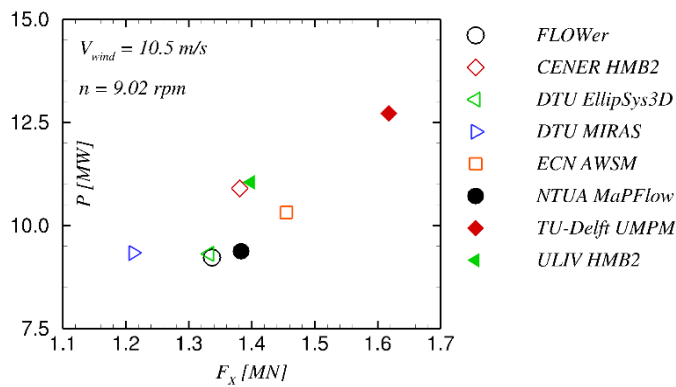

Fig. 5 Comparison of power and thrust with the other project partners [9]

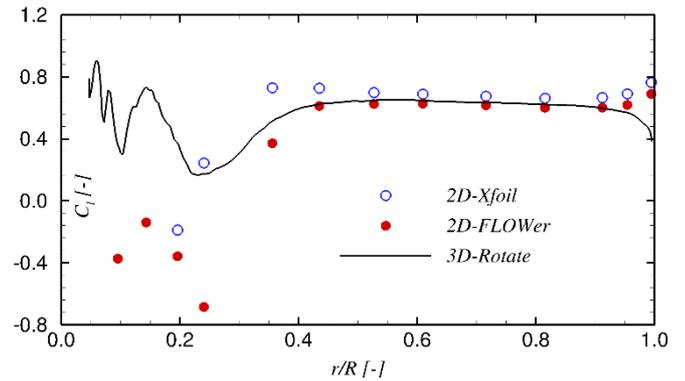

Fig. 6 Averaged 3D and 2D lift coefficients.

*C. Rotational Effects*

The comparison of the 3D and 2D calculations are presented in Fig. 6. The *XFOIL* results are also presented for comparison. The lift coefficient ($C_l$) of the 3D case shows remarkably higher value at the inner part of the blade, and the augmentation becomes smaller as the radial position increases. The enhancement of $C_l$ is observed until $r/R = 0.4$. The value of 2D $C_l$ predicted by *FLOWer* at the middle part of the blade is equal with the 3D case. This is consistent with the other literatures findings [1-3,7], stating that the rotational effects have small or no influence on the flow development at the middle part of the blade (at the attached flow region). At the proximity of the blade tip, the presence of trailing vortices induces a downwash, which in turn decreases $C_l$ as compared with the 2D case.

The *XFOIL* calculations consistently show higher values compared with the flower calculations. Two main reasons can be used to explain the results. First, the AVATAR blade were constructed from relatively thick airfoils (more than 20% maximum thickness), resulting in a higher adverse pressure gradient, which makes the prediction by means of boundary layer codes such as *XFOIL* become less accurate. The second reason is related to transition points. The *XFOIL* code has no ability to set the fully turbulent inflow condition; therefore, the transition prescription method by means of tripping has to be used. The present investigation prescribed the transition points at 5% chord on the suction and pressure sides. This might have an influence on the results since the thick airfoil is more sensitive to the location of transition point, i.e. high impact on the arc length of the laminar flow.

It can be observed that the trend of 3D $C_l$ over the radius is similar with the corresponding 2D calculations (*FLOWer* and *XFOIL*). The augmentations seem to increase as the radial position decreases; however, it still follows the shape of the lift curve from the 2D case. One may conclude that the rotational effects are airfoil dependent - apart from the dependency of rotational effects on the local solidity ($c/r$) as mentioned in Ref. [4,5,19].

Fig. 7 shows the averaged flow field of the non-dimensional streamwise velocity ($u/U$), variable $u$ represents the streamwise velocity and variable $U$ represents the wind speed (10.5 m/s). It is clearly shown that the 3D case produces delayed separation and smaller wake size. This is believed as the work of the centrifugal and Coriolis forces. The centrifugal force transports the separated flow at the near hub region towards the middle part of the blade, as shown in Fig. 4. A significant reduction in the wake size is then created, which results in the boundary layer thinning. This increases the fluid velocity on the suction side of the blade, which in turn reduces the pressure level. The Coriolis force acts in the streamwise direction. It delays the point of separation and contributes to $C_l$ increase in Fig. 6.

The present results agree with Sicot *et al.* [3] who observed that the rotation decreases the pressure level on the suction side of the blade, which then increases the lift coefficient. However, his conclusion regarding separation point contradicted the present results and previous literatures, for instance, Du and Selig [1] and Herráez *et al.* [2]. It was shown that separation is delayed in the 3D case as compared with the non-rotating 2D case, and this behaviour didn't occurred in the Sicot's results.

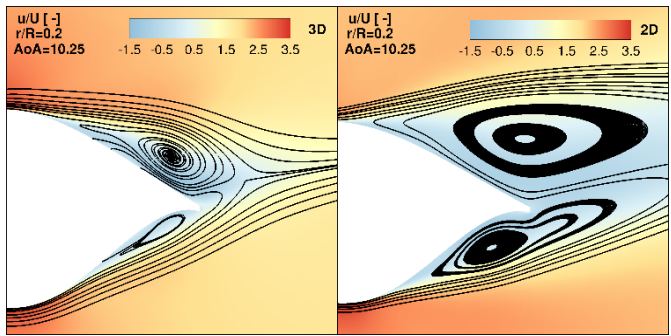

Fig. 7 Averaged flow field for 3D (left) and 2D (right) cases.

### IV. CONCLUSION

An investigation of rotational effects on large wind turbine blades has been conducted. The rotation plays an important role only at the inner part of the blade. It was observed that $C_l$ augmentation is airfoil dependent. A massive reduction of the wake size and separation delay are observed at the inboard blade section. The radial flow development at the inner part of the blade creates a smaller wake size. This reduces the pressure level on the suction side, which in turn increases the lift coefficient at the corresponding blade section.

### ACKNOWLEDGEMENTS

The authors gratefully acknowledge the High Performance Computing Center Stuttgart (HLRS) for the computational resources on the LAKI cluster, the AVATAR project for the provision of the blade and the validation data, the Indonesian Directorate General of Higher Education (DIKTI) for the scholarship grant, and the members of Aircraft Aerodynamics and Wind Energy research group at the IAG for the valuable scripts and discussions.